# Verifying Design through Generative Visualization of Neural Activities

**Pan Wang[1], Danlin Peng[1], Simiao Yu[1], Chao Wu[3], Peter Childs[1], Yike Guo[1,4] and Ling Li[2]**
[1]Imperial College London, United Kingdom
[2]University of Kent, United Kingdom
[3]Zhejiang University, China
[4]Hong Kong Baptist University

## ABSTRACT

Understanding human cognition plays a vital role in verifying the effectiveness of a design. Great progress has been made in the past in order to provide such understanding in cognitive neuroscience. However, current neuroscience-focused approaches for evaluating design are limited due to the lack of direct visualization of the mental activities. Seeking a tool to visualize the states of the mind from the measured brain signals/images is an intriguing challenge in interpreting cognition and also in understanding design impact. To tackle this challenge, we are inspired by the work of S.Palazzo et al. who introduced a mental image reconstruction method through measured electroencephalogram (EEG) using a generative adversarial network (GAN). Based on his work, we proposed a framework of revealing design impact to the brain by reconstructing mental images representing what is emerged in the brain when a design is presented. First, a recurrent neural network is used as the encoder to learn a latent representation from the raw EEG signals, which were recorded while subjects were looking at 50 categories of images. Then, a generative adversarial network conditioned on the EEG latent representation is trained for reconstructing these images. After training, the neural network is able to reconstruct mental images from brain activity recordings. To demonstrate the proposed method in the context of design verification, we performed a case study, in which we presented a set of iconic design images sequentially to the subject to explore if a subject had created a cognitive association. Each subject's brain activities related with a design image were recorded and fed to the proposed image reconstruction model to generate mental images. The experimental results indicate that a successful design could inspire the subject to associate the design with ideas or valued products. For instance, when subjects were shown an image of a bitten apple, a mental image of a phone instead of the apple itself was reconstructed, illustrating the cognitive association with the brand icon. The proposed method could have a great potential in verifying iconic designs by visualizing the cognitive understanding of the underlying human brain activities.

## CCS CONCEPTS

• Cognitive association → Mental image reconstruction; Deep learning; Visualization

## KEYWORDS

Cognitive association, Deep learning, Mental image reconstruction, Brain decoding, EEG

## 1 INTRODUCTION

We build our conscious awareness through senses and perceptions. However, what we see may not be what we think. Human has a distinguished ability of creating cognitive association. For example, a stimuli received from our sensory system may involve an associated cognitive pattern in the brain which can be quite different from its original form and the stimuli itself. The purpose of brand design in general is to take advantage of this ability to establish a cognitive association between a brand icon with the corresponding product for consumers. Therefore, being able to understand and verify this relationship is crucial. Establishing and enhancing such a cognitive association is exactly what a branding excise about. In fact, the research in cognitive association already has a far-reaching history, such as Pavlov's research and some research in experimental psychology have shown that a response usually elicited by the potent stimulus in the experiment [24][21][1][22]. However, there is still lacking of understanding in this key cognitive function. It would be ideal of we can visualize directly what is in the brain to see if it is exactly the associated product when a design icon is represented to a subject. Such a "mind reading" technology could have a profound impact to the design industry. In Neuroscience, such a "mind reading" technology has been developed under the name of "brain decoding" [16][7][10] [27], where the brain activity is predicted via a model constructed from brain imaging, measured with imaging technologies, such as electroencephalography (EEG) or functional magnetic resonance imaging (fMRI), labelled with stimuli (see Fig 1). This method aims to create a model to map between the brain activation pattern with the image itself, as illustrated in Figure 1. Recently, some deep learning based brain decoding models [20][25][8][6][10][23] have been proposed to reconstruct subjects' mental image from their brain activities, measured with EEG[15] or fMRI[14], when a real image is presented. These methods take a generative approach so that the decoding is achieved by generating an image representation of the thinking in the brain. This is achieved through a generative model training to link the features of images with those of brain imaging. We recognizes that this generative approach on brain decoding provides a promising way to studying cognitive association, since the generative model can be trained through learning representations of real world images and brain imaging as well as their relations. Thus, it is possible to take this generative approach to build a framework to generate cognitive associations for visualization when being presenting a stimuli.

Neuroscience-inspired design.

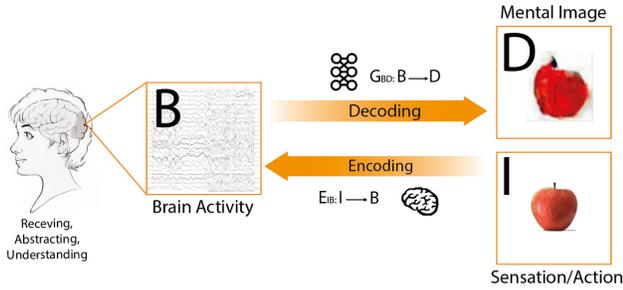

**Figure 1: Traditional brain encoding/decoding method**

In this work, we attempt to build such a framework to visualize human mental association in the context of verifying the impact of brand design. We have designed an enhanced generative model under generative adversarial network with auxiliary classifier framework to map brain activity to mental associations.

Given the development requires a sophisticated experimental design for collecting quality data, we present the development of the framework in terms of EEG analysis. We conducted two sets of experiments, the image presentation experiment for building the models and mental association experiment for inference to generate mental images illustrating the underlying mental associations. As illustrated in Figure 2, in the image presentation session, we established an encoder which can map a brain signal $B$ to a mental image $D$. Subjects are required to view a set of images $D'$ without any associative thinking, which ensures that the mental images $D$ are same as the seen images $D'$. We then train an encoder to project brain signal $B$ to the latent space $Z_B$ which is equivalent to $Z_D$ that encodes the representative features of the mental image $D$. After this, an image generator trained within a conditional GAN framework with an auxiliary classifier is to discover the intrinsic relation between the representations $Z_B$ and $Z_D$, thereby establishing the path for visualizing mental images using the brain signal.

Secondly, in the mental association session, we explore a human's real cognitive understanding of designs. We record the brain signal when subjects see a design and then employ the trained encoder to extract the class discriminative EEG feature vector representing the mental image concept. After this, the trained image generator is used to visualize the mental image from the EEG feature vector.

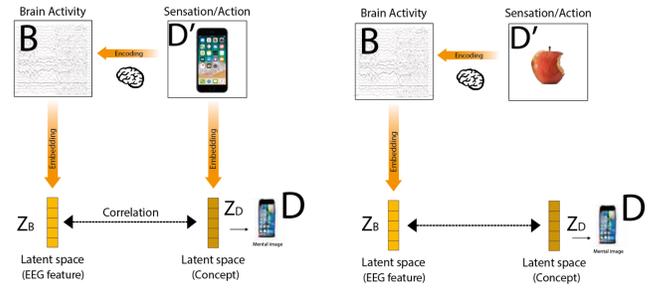

**Figure 2: The representational paradigm**

The main contributions of this paper consist of three parts as follows:
- We proposed and developed a general framework of design verification with the generative brain decoding approach where verification can be done by visualizing cognitive association with the seen images by generating mental images.
- The experimental results show that our model can reconstruct mental imagery images (i.e. the cognitive representation) with high quality. A case study is performed by testing five iconic designs and reconstructing their corresponding mental images. The results show that subjects associate a normal apple with an apple image and relate a bitten apple to the "Apple" brand product.
- We visualize the mental process by generating the temporal dynamic changes of mental images inferred from the time sequence EEG data, which provides a way to understand the process of a cognitive transformation of while a subject receives a stimulus.

We believe that our work will have a broad applications in many fields such as neuromarketing, cognitive science and in design industry.

## 2 RELATED WORK

Our work can be generally categorized as brain decoding since we reveal the contents of the brain's cognitive activity. Traditional brain decoding predicts the information contained in brain neural activities by utilizing an encoding model trained with brain imaging data labelled with stimuli [16], and studies have mainly focused on linear classification. The brain activity and visual contents have been coded through both univariate and multivariate mapping[17][28][18] which provide a better understanding between encoding and decoding analysis. Brain decoding for understanding cognition from brain activities have involved different statistic models to define the cognitive stage[12]. Wegrzyn et al.[29] have explored the cognitive process and make predictions regarding the language in cognitive domain. The research has resulted in a 'keyword space' derived from the NeuroSynth database. Brain-computer interface approaches focus on visualizing the collinear pattern of brain to predict the mental

state[13][30], which also do not have a direct visualization. All these methods are based on the predictive approach. Thus, the decoded content has to be in the set of labels of training data in building the corresponding encoding system. With the recent development of deep learning, an approach[2] was proposed for modelling cognitive events from multi-channel EEG time-series by transforming EEG activities into topology preserving multispectral images, then train a deep recurrent convolutional network to do the classification. Most previous studies have also mainly focused on a classification based approach. For instance, training a classifier based on topographic maps generated by EEG signals[11], to learn a relationship between brain signal and the simple executive commands or seen image classes[25]; Identify functional brain networks and use network similarity algorithm to distinguish between categories[9]. One of the most recent developments of brain decoding is based on the generative model approach where brain decoding is done by a generative model learned through associating features of brain imaging measurement of neural activities with those of stimuli. Instead of only being able to decode the simple content classes, something more inspiring and more complex results and seen images can be reconstructed directly based on this generative approach thank to the associating features of brain imaging and stimuli at the representation level. Palazzo proposed a GAN (Generative Adversarial Nets) based brain encoding framework using the EEG signals[20]. Reconstructing an imaginary visual image from EEG activity has also been proposed by Tirupattur et al. [26] use of a deep learning method to reconstruct the visual image from fMRI brain activity has been explored by[8]. These methods give the potential to reconstruct a cognitive mental image from human brain activity under different stimuli.

## 3 A MENTAL IMAGE VISUALIZATION METHOD AND CASE STUDIES

### 3.1 Problem Formulation and method overview.

The primary goal of this work is to visualize the mental image using brain activities thereby interpreting human cognition association to assess the effectiveness of the design. As shown in Figure 2, the problem can be formulated as follows. Given a target seen image domain $D'$ and a mental image domain $D$ containing the real cognitive understanding of the corresponding seen image, we denote $Z_D$ as the shared latent space which contains the cognitive representations of domain $D'$ and $D$. We also have a brain signal domain $B$ containing EEG signal recorded while subject seeing or imagining images and we attempt to encode $B$ with the corresponding latent space $Z_B$. We then make three key assumptions of the mental image visualization problem: 1) The mental images are as same as the seen image if a subject is only presented an image in a very short time and ask them to focus on the image without any associative thinking. 2) A meaningful low-dimensional class discriminative representation of what we call EEG features $\{z_k\}_{k=1}^N \in Z_B$ can be learned from multidimensional and highly-noisy raw EEG data. 3) For the same image, the visually-relevant EEG features and the imaginarily-relevant EEG features have some intrinsic relation.

Based on these assumptions, the objective of mental image visualization problem is to learn a generating function $G_{BD}: B \rightarrow D$. Since EEG data is a time sequence signal, it is difficult to learn the direct mapping relation between mental image and brain signal directly. Nonetheless, we could still learn it in an implicit fashion using deep learning method. The model training consists of two phases which is shown in Figure 3.: 1) Training an EEG feature encoder to embed raw EEG data $\{b_k\}_{k=1}^N \in B$ into a latent representation $\{z_k\}_{k=1}^N \in Z_b$. 2) Training an image generator under a conditional GAN (ACGAN) [19]framework to discover the intrinsic relation between EEG feature $z_k$ and corresponding mental image $d_k$ which is close to the image stimuli $d'_k$ in presentation experiment, thereby visualizing mental image using brain signal.

More specifically, in the first stage, we record the EEG signal while subjects are viewing the visual image stimuli and train a LSTM-based[3] encoder to extract class–discriminative EEG features from the raw EEG data. The extracted EEG features and corresponding image stimuli are then used to train the ACGAN model as an image generator condition to the feature of the EEG signal. Finally, we use the trained encoder to extract meaningful EEG features from the EEG data which are recorded while subjects are presented an iconic image with the requirement to think its meaning and then feed the extracted EEG feature into the well-trained GAN model to generate the mental imagery image.

The results show that an iconic design has a huge influence on the human's cognitive understanding. When people are shown some famous icons and instructed to imagine something, most of them will imagine relevant products rather than the logo itself. Companies can use our model to assess the effectiveness of a design or advertisement.

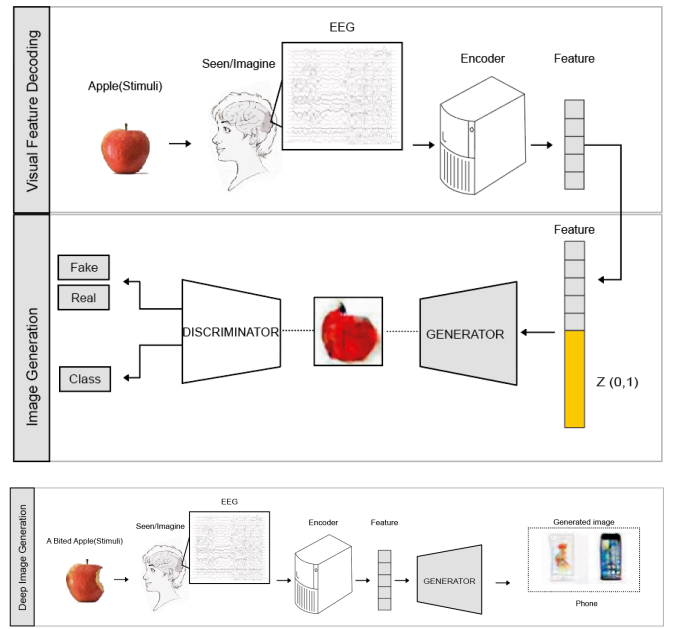

**Figure 3: Overview of the proposed deep learning architecture for mental image generation from EEG signal.**

## 3.2 Building EEG feature encoder

The goal of the encoder is to extract meaningful representations from the multichannel temporal EEG sequence. In order to track time dependencies in the EEG data, LSTM recurrent neural network was employed to interpret temporal dynamics and extract EEG features. As shown in Fig. 11, the EEG feature encoder is made up of a standard LSTM layer and followed by two fully-connected layers (FCN). ReLU nonlinearity is added after each fully-connected layer. A Softmax layer is appended as a classification module. During training, we feed the data of all EEG channels at each time step into the LSTM layer. The output of the LSTM layer at the last time step is used as the input of the fully-connected layer and the final output is regarded as the EEG feature vector which represents the class-discriminative brain activity information. The dataset is split into 3 sets: 80% (2,000 images) for training, 10% (250) for validation, 10% (250) for testing. The learning rate is initialized to 0.0001 and Adam gradient descent method with batches of size 16 is used to learn the model parameters. Hyperparameters are tuned on the validation set.

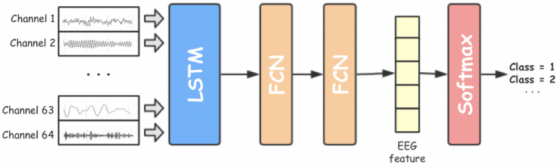

**Figure 4: EEG feature encoder**

## 3.3 Building Image Generator

In our multi-class scenario, given class labels, the generator must be capable of generating correct images from 50 classes. However, the original GAN[5] framework is not able to deal with multi-class tasks, and since the generator only utilizes the simple fully connected neural network which is hard to learn proper features from complicated images to generate high-quality images. Therefore, we train our image reconstruction model under the 'Deep Convolutional Generative Adversarial with Auxiliary Classifier' framework which not only can generate images from specified classes but also has the ability to generate satisfying images through capturing the data distribution of complicated nature images. Figure 5 shows the high-level view of the whole architecture of the Generative Adversarial Network that we implement in the experiment.

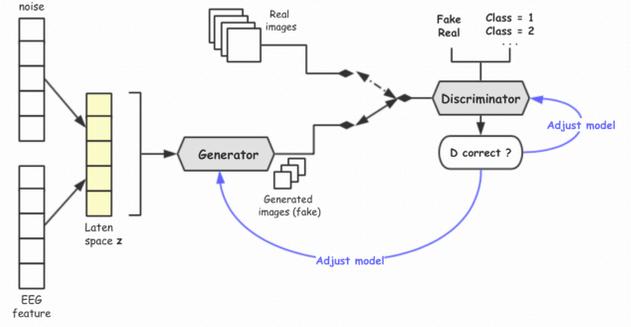

**Figure 5 : General view on architecture**

*ACGAN.* The foundation of this framework is ACGAN which consists of two parts: a generative model $G$ and two discriminative models $D_a$ and $D_b$. The generator $G(z|c)$ is trained to capture the target data distribution $p_{data}(x)$ from the condition EEG feature $c$ of class $y$ and noise distribution $p_z(z)$, and aim to generate images of the target class as real as possible to make the discriminator recognize the generated images are real. Whereas the discriminative model $D_a(x|y)$ is a binary class classifier which distinguishes whether a sample image belongs to the real image set. The discriminative model $D_b(x|y)$ is a muti-class classifier which identify the image class. Both the generative and discriminative models are trained simultaneously and play against with each other to minimax the log likelihood value function V (D, G).

$$\min_G \max_D V(D,G) = \mathbb{E}_{x \in p_{data}(x)}[log D_a(x|y) + log D_b(x|y)] \\ + \mathbb{E}_{z \in p_z(z)}[log\,log(1 - D_a(G(x|c)|y)) \\ + log D_b(G(x|c)|y)]$$

## 3.4 The structure of the Generator

Inspired by ACGAN, we employ a generator with 5 upsampling layers. The overview of generator architecture is shown in Figure 6.

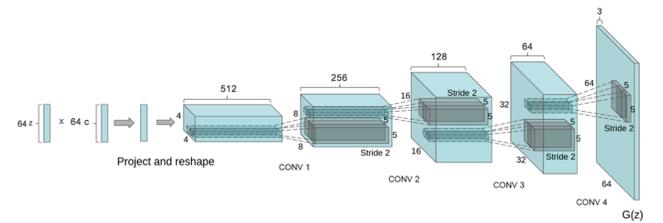

**Figure 6: Architecture of the generator**

The generator takes as input a high level EEG representation which is the matrix multiplication result of 64-dimensional random Gaussian noise and 64-dimensional EEG features. This input is then fed into the first transposed convolutional layers, which spatially upsamples the input vector by four times and output 512 feature maps. The output of the first layer still needs to go through another four transposed convolutional layers where the feature map size is doubled and the number of features map is halved after each layer so that the final output image size is 64*64 pixel with three color channels. Batch normalisation and LeakyReLU nonlinearities are added after each transposed convolutional layer. Table 1. shows the detailed architecture of the generator and the main parameters value we choose for each layer.

Table 1: Generator architecture

| Operation | Kernel | Strides | Padding | Feature maps | Map size |
|---|---|---|---|---|---|
| Transposed Convolution<br>Batch Normalization<br>Activation-LeakyReLU | 4*4 | 1 | 0 | 512 | 4*4 |
| Transposed Convolution<br>Batch Normalization<br>Activation-LeakyReLU | 4*4 | 2 | 1 | 256 | 8*8 |
| Transposed Convolution<br>Batch Normalization<br>Activation-LeakyReLU | 4*4 | 2 | 1 | 128 | 16*16 |
| Transposed Convolution<br>Batch Normalization<br>Activation-LeakyReLU | 4*4 | 2 | 1 | 64 | 32*32 |
| Transposed Convolution<br>Activation-tanh | 4*4 | 2 | 1 | 3 | 64*64 |

### 3.5 The Structure of the Discriminator

Inspired by VGG16[4], Our discriminator has two main modules: one is the convolutional module, mainly used extract the primary feature from the images. The other is the classification module including two classifiers. One of the classifiers in the classification module is a binary classifier to distinguish between fake and real images, and the other one is a multi-classifier which is used to recognize the image class.

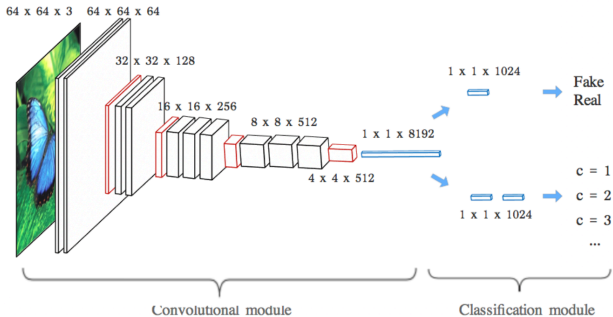

Figure 7: Architecture of the discriminator

*Convolutional module:* The convolutional part of the discriminator is made up of 10 convolutional layers. Batch normalisation and LeakyReLU nonlinearities are added after each convolutional layer. This takes as input 64x64 pixel images with three color channels. The number of feature maps starts at 64 at the first layer and is doubled at layer three, five and eight respectively, after that the number of feature maps reaches 512 at the last convolutional layer. Feature map size remains the same at each convolutional layer, whereas the max pooling layer halves the feature map size. The input goes through each convolutional layer, and the feature map size is halved at the max-pooling step which is added after the second, the fourth, the seventh and the tenth convolutional layers. The feature map size is 4x4 after the final layer of the convolutional module.

*Classification module.* The output data size of the convolutional module we mentioned before is $4 \times 4 \times 512$. We flatten such output data and fed it into the following two classifiers to distinguish real images and recognize the image class

*Binary Classifier.* The binary classifier aims to calculate the probability that an image comes from the real image dataset rather than the generative model. It consists of two fully connected layers to reduce the number of features to 1024 and 1. We use ReLU as the activation function after the first fully-connected layer and add a sigmoidal probability estimate after the second fully-connected layer.

*Multi-class classifier.* The multi-class classifier is designed to predict which class the input image belongs to. It consists of three fully-connected layers. The first layer reduces the number of features to 1024 followed by a hidden layer. After this, the data is fed into the last fully-connected layer where the number of features is reduced to the number of image categories. We use ReLU as the activation function after the first and the second fully-connected layer and add the Softmax layer after the last fully-connected layer.

### 3.6 System performance evaluation

*3.6.1 Evaluation for encoder.* We employ classification rate to evaluate the performance. As shown in Table 2, the accuracy for the whole dataset which contains 50 classes is 70.8%.

| Model | Validation set | Test set |
|---|---|---|
| LSTM + nonlinear | 71.20% | 70.80% |

Table 2: Classification rate on validation set and test set

*3.6.2 Evaluation for generator*

**Confusion matrix.** We generate 1500 image per class and classify them by the classifier which is trained on real image dataset. Subsequently, we obtain the confusion matrix as shown in Figure 12, and calculate the corresponding evaluation indices such as precision, recall and F1 score. As we can see in the table, the Phone has the highest F1 score which means the generated image has similar features to the real image so that it can be easily recognised by the classifier. However, Coke has the lowest F1 score which means the generator has difficulty in capturing the object features for some reason. We find that under the ACGAN framework, the generator tends to produce those easy-to-classify images with high-quality and stay away from hard-to-classify images.

**Inception score-generator.** Another method to estimate the realism and diversity of produced images is the Inception score(Salimans et al. 2016), and this correlates well with human evaluation. We can obtain the conditional label distribution $p(y|x)$ using the Inception model. For the images that contain meaningful items, the distribution $p(y|x)$ should have low entropy. If the model can generate diverse images the marginal $\int (y|x = G(z))dzp$ should have high entropy. By combining these two scores, we can obtain the final inception score as $exp(E_xKL(p(y|x)||p(y)))$ (we exponentiate the score in practice).

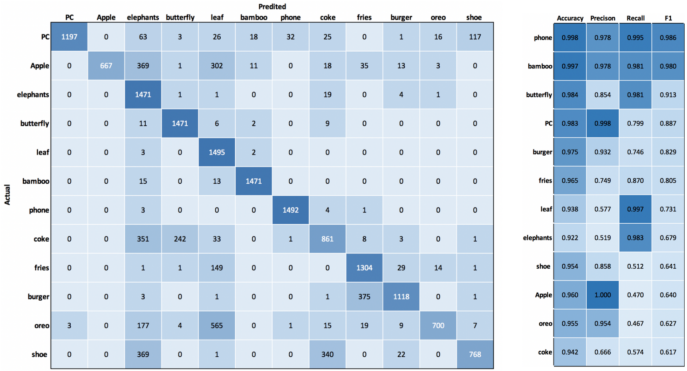

Figure 8: Confusion matrix for VAC+GAN method

Table 3: Inception score

| Class | Inception Score |
|---|---|
| phone | 6.6 |
| bamboo | 5.4 |
| butterfly | 5.2 |
| PC | 6.0 |
| burger | 6.8 |
| fries | 5.9 |
| leaf | 5.5 |
| elephants | 7.3 |
| shoe | 6.2 |
| Apple | 6.9 |
| oreo | 7.5 |
| coke | 6.1 |

## 4 Experiment Design.

The experimental programme was conducted in two parts, an image presentation experiment for collecting training data and a mental imagery experiment for cognitive association reconstruction.

### 4.1 presentation experiments

In the image presentation experiment, the stimuli contains 50 categories, 50 images in each category. Each run contains 1 repeated image which was used to maintain their attention of the presented images by pressing a button. We designed a fixation spot at the beginning, which was presented in the central of the images with its color changed from black to red for 0.5 s. At the beginning and end of each run, 10 s were added as a rest time. Each image was presented 0.75s, and the subject could stop the presentation at any time during the experiment.

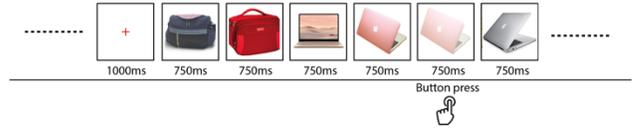

Figure 9: Image presentation experiment. In the image presentation experiment, images were presented in the middle of the screen with the same size, same length to subjects and a central fixation point.

### 4.2 Imagery experiment.

In the imagery experiment, subjects were required to visually imagine the associated objects or design inspirations from the presented images. The whole experiment consisted of 6 blocks, each block consisting of 10 imagery trials. Each imagery block consisted of a 4000 ms cue period, an 8000 ms stimuli display period, a 15000 ms drawing period, an 8000 ms imagery period and 3000 ms evaluation period. A red cross presented in the centre of the display worked as a fixation trail. In addition, we added a 5000 ms rest period at the beginning and end of each block. Subjects were required to associate required iconic logo immediately after the cue period, then drawing it downing during the drawing period. After a beep, we used an 8000 ms imagery period, followed by a 3000ms evaluation period in which required the subjects to evaluate the correctness and vividness of their imagination on '1-5' five levels(5-very vivid, 4-fairly vivid, 3-rather vivid, 2-not vivid, 1-cannot correctly recognize the target)by writing down the numbers in the form. The subject could stop at any time during the experiment.

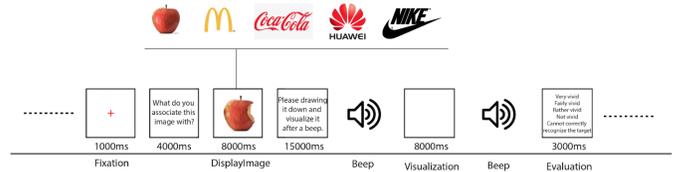

Figure 10: Imagery experiment design. In the imagery experiment, subjects were required to visualize imagery image when then seen different stimulus followed by experiment design.

### 4.3 Subjects and Equipment.

Six healthy right-handed subjects (three males and three males) aged 18-30 years with normal or corrected-to-normal vision participated in this experiment. The participants were all native English speakers. All subjects had considerable training in EEG experiments, and the study was approved by the Imperial College Research Ethics Committee (ethics statement, ICREC reference: 18IC4848). We used the Neuroscan electroencephalogram (EEG) equipment, and the signal was recorded (band-pass 0.05– 100 Hz, sampling rate 500 Hz) from a set of 64 Ag/AgCl electrodes according to the 10–20 system with the Neuroscan Synamp2 Amplifier (Scan 4.3.1, Neurosoft Labs, Inc. Virginia, USA). The room was electrically shielded, dimly lit and sound attenuated with a comfortable seating. The EEG signals were filtered in run-time with frequency boundaries 14-70 Hz which contains the necessary bans(Alpha, Beta, and Gamma) that are most meaningful for the visual recognition and imagination. The sampling frequency was set to 1000 Hz and the quantization resolution to 16 bit. Data was collected from each subject over several runs to minimize the confounding factors.

### 4.4 Visual stimuli.

The subjects were shown 50 images from 50 different categories images for a total of 2500 images each person in the image presentation experiment. 5 icons (Apple, Coca-cola, Macdonald, Huawei and Oreo). The stimuli images were cropped to the center and were resized to 500 × 500 pixels.

## 5 Results.

As mentioned previously, we trained an encoder to extract features from raw EEG data and also trained a generative model under ACGAN framework to reconstruct seen images. The results showed that we could employ this well-trained encoder and generator as the cognitive translator, translating the cognitive image from the human brain. Our results consist of three parts: Firstly, we introduce the seen image reconstruction result, and then we discuss how subjects interpret the design semantics by reconstructing the imagined object. Finally, we describe how cognitive images change at different imaginary time points.

### 5.1 Seen image reconstruction

Seen image reconstruction is the baseline of our work. As shown in Figure 4, the ACGAN model discovered the intrinsic relation between brain signal and seen image and successfully reconstructs the correlated mental image.

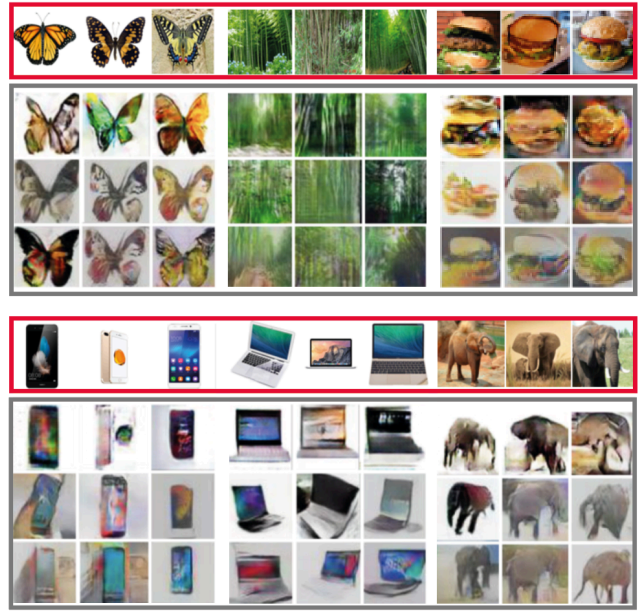

**Figure 11:** Images show the presented and generated images from image presentation in each class. The images in the red frame are the seen images, the images in the grey frame are the reconstructed images from the image presentation experiment.

### 5.2 Iconic brand experiment result

The results of the mental imagery experiment showed that the image reconstruction model is able to generate the mental association from brain activity, which presents an effective method for reading a human's mind and interpreting human cognition.

The aim of this experiment is to estimate the effectiveness of the design semantics by observing how mental images change when the subjects saw an object with different design semantics. In the apple or Apple logo session, we presented an image of an apple and an image of an apple with a bite removed to the subjects and examined if the subjects will imagine the apple itself or some other related objects.

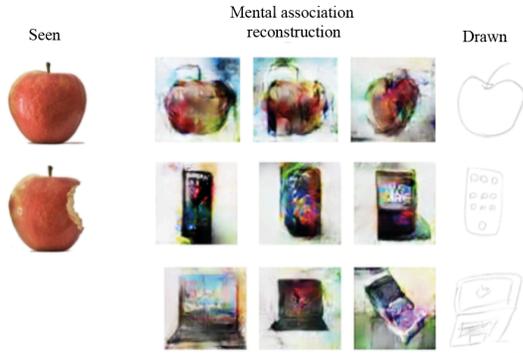

**Figure 12: Cognitive image for apple and apple with a bit**

As shown in Figure 12, when the subject saw the apple, most of them associated the apple itself; whereas when they saw the same apple with a bite removed, they immediately associated the cue image with some relevant products such as a phone or a laptop. This showed the subjects regarded the bitten apple as the Apple logo, which verifies how a slight change in the design semantics may have a huge influence on a human's cognition.

Similarly, as indicated in Figure 13 we presented the logos of Mcdonald's, Nike and Coca-Cola to subjects. Most of the subjects imagined the products related to the logo. When the subjects saw the Mcdonald's logo, they imagined fries, burger and coke. Subjects imagined shoes when saw the Nike logo, and imagined coke when saw the Coca-Cola logo.

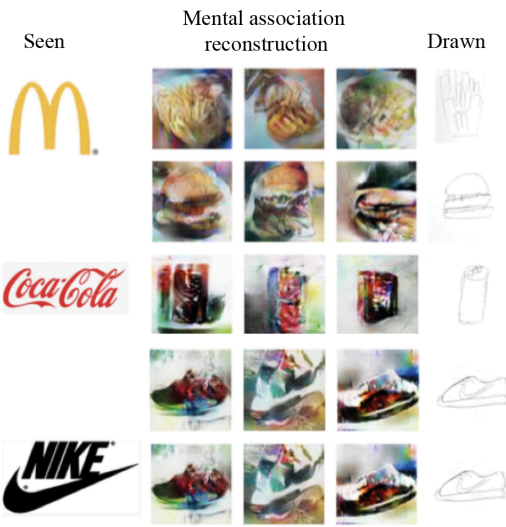

**Figure 13: Cognitive association images for the logos of Mcdonalds, Nike and Coca-Cola.**

## 5.3 Cognitive associations at different time point

In order to have a deep understanding of how people relate to these cognitive images, we explored the changes of the cognitive associations in different time.

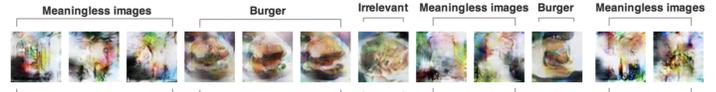

**Figure 14: A sequence of generated cognitive images**

Figure 14 demonstrated a sequence of generated mental images that show how the imagined content changes in a 1 second period. There were three types of cognitive images: target class images, irrelevant class images and meaningless images.

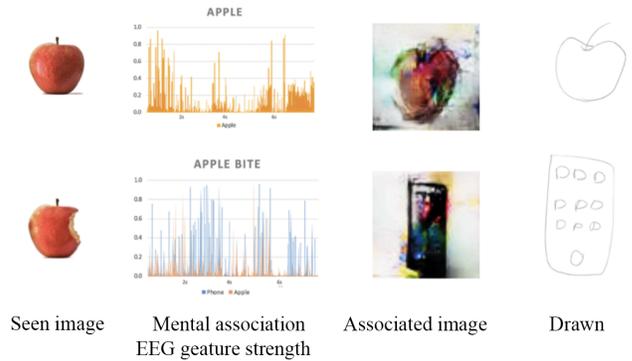

Seen image | Mental association EEG geature strength | Associated image | Drawn

**Figure 15: Time-related changes in cognition: apple and apple with a bite: The first column shows the presented cue stimuli; The third column shows the cognitive image; The fourth column shows the corresponding drawings from the subject.**

For each 8-second mental imagery session, around 30%-70% brain signal is regarded as meaningless, 10%-50% can be recognized as belonging to the target class. As shown in the lower part of the second column, we find that after seeing an apple with a bite, the subject is mainly imagining phones instead of the apple itself.

## 6 Conclusion and Discussion

In this research, we have successfully verified that the cognitive mental image can be visualized by using our image reconstruction model conditioned by EEG features. By using the deep neural network which translates the cognitive image from the human brain in our experiment design, we find the real cognitive image which is the mental image, could be visualized by the image reconstruction model, which is not as same as the visual image. The experimental results also indicate that a successful design could inspire the subject to associates the design with ideas or valued products. Therefore we propose a new method of verifying the design by visualizing the cognitive understanding from the human brain and demonstrate a new way of evaluating the effectiveness of the design semantic.

In addition, generating the temporal dynamic changes of mental images inferred from the time sequence EEG data has been used to understand the cognitive transformation. This could potentially be used in other purposes such as exploring the cognitive process from brain activation. In the future, improvements can be made in increasing the number of image categories which is expected to enable our model to reconstruct more features from different images.

## ACKNOWLEDGMENTS

The authors would like to acknowledge Zhejiang University Neuromanagement lab for providing equipment and subject recruitment to facilitate the necessary environment for EEG data acquisition.